\documentclass[twocolumn,english,prl]{revtex4}
\usepackage{ae,aecompl}
\usepackage{helvet}
\usepackage[T1]{fontenc}
\usepackage[latin9]{inputenc}
\usepackage{babel}
\usepackage{amsmath}
\usepackage{amssymb}
\usepackage{graphicx}
\usepackage{esint}
\usepackage[unicode=true,pdfusetitle,
 bookmarks=true,bookmarksnumbered=false,bookmarksopen=false,
 breaklinks=false,pdfborder={0 0 1},backref=false,colorlinks=false]
 {hyperref}

\makeatletter
\@ifundefined{textcolor}{}
{%
 \definecolor{BLACK}{gray}{0}
 \definecolor{WHITE}{gray}{1}
 \definecolor{RED}{rgb}{1,0,0}
 \definecolor{GREEN}{rgb}{0,1,0}
 \definecolor{BLUE}{rgb}{0,0,1}
 \definecolor{CYAN}{cmyk}{1,0,0,0}
 \definecolor{MAGENTA}{cmyk}{0,1,0,0}
 \definecolor{YELLOW}{cmyk}{0,0,1,0}
}

\makeatother

\begin{document}

\title{Internal loss of superconducting resonators induced by interacting
two level systems}

\author{Lara Faoro$^{1,2}$ and Lev B. Ioffe$^{2}$}

\affiliation{$^{1}$ Laboratoire de Physique Theorique et Hautes Energies, CNRS
UMR 7589, Universites Paris 6 et 7, 4 place Jussieu, 75252 Paris,
Cedex 05, France}

\affiliation{$^{2}$ Department of Physics and Astronomy,~Rutgers The State University
of New Jersey, 136 Frelinghuysen Rd, Piscataway,~08854 New Jersey,
USA }

\date{\today }
\begin{abstract}
In a number of recent experiments with microwave high quality superconducting
coplanar waveguide (CPW) resonators an anomalously weak power dependence
of the quality factor has been observed. We argue that this observation
implies that the monochromatic radiation does not saturate the Two
Level Systems (TLS) located at the interface oxide surfaces of the
resonator and suggests the importance of their interactions. We estimate
the microwave loss due to $\mathit{interacting}$ TLS and show that
the interactions between TLS lead to a drift of their energies that
result in a much slower, logarithmic dependence of their absorption
on the radiation power in agreement with the data. 
\end{abstract}


\maketitle
High quality superconducting CPW resonators are used in a number of
diverse fields, ranging from astronomical photon detection \citep{Klapwijk2011,Day2003}
to circuit quantum electrodynamics \citep{Wallraff2004,DiCarlo2009,Ansmann2009,Steffen2010}.
In these applications, the CPW resonator is operated in a regime of
low temperature $(\sim10$mk) and low excitation power (single photon).
The performance of these devices is directly related to the resonator
quality factor, $Q$, defined by the photon decay rate, $\gamma_{ph},$
as $Q=\omega_{0}/\gamma_{ph}$, where $\omega_{0}$ is the resonance
frequency. $\gamma_{ph}$ is given by the sum of the escape rate from
the resonator and the rate of the intrinsic decay; the latter sets
the limit on the performance. At low temperature and single photon
regime, the intrinsic decay is usually attributed to the excitations
of TLS located in the dielectrics (bulk and surfaces) surrounding
the resonator \citep{Martinis2005}. This belief is strongly supported
by the observation of temperature-dependent resonance frequency shift
that closely agrees with the one predicted by the conventional theory
of microwave absorption of TLS in glasses \citep{Anderson1972,Black1977}.
According to this theory, one expects also that, as the power of the
radiation applied to resonator is increased, TLS in the dielectrics
get saturated, thereby limiting the maximal power that can be dissipated
by photons. This results in a strong power dependence of the quality
factor: $Q\propto\sqrt{P}$ above a critical power $P_{c}$.
This power
dependence is indeed observed in many resonators characterized by intrinsic loss tangent ${\sim 10^{-3}}$ at very low powers \citep{Lindstrom2009,Pappas2011a}.
However resonators characterized by lower intrinsic loss at low powers typically show much weaker power
dependence \citep{Wang2009,Macha2010,Wisbey2010,Khalil2011,Sage2011}.

In fact, careful fitting of the loss versus power to the empirical
equation ${\displaystyle Q\propto(1+P/P_{c})^{\varphi}}$ gives $\varphi\sim0.03-0.16$
for the resonators made of Nb and Al on sapphire and of Nb on undoped
silicon \citep{Macha2010}; similarly high-Q single-layer resonators
of different geometries made of Al on sapphire substrate show $\varphi\sim0.1-0.2$
\citep{Khalil2011}. Corrections of the conventional theory of TLS
which include the geometric dependence of the applied electric field
failed to reproduce the weak power dependence observed experimentally
\citep{Wang2009,Wisbey2010,Khalil2011,Sage2011}.

The failure of the conventional theory of TLS to predict the power
dependence of the quality factor for the high quality resonators is
an indication of a serious gap in our understanding of TLS in amorphous
insulators. In this Letter we argue that in high-Q superconducting
CPW resonators the TLS located in the interface oxide surfaces are
subject to stronger interactions than the TLS located in the bulk
dielectrics. These TLS interactions lead to a drift of the TLS energies
that results in a logarithmic dependence of their absorption on the
radiation power, $P$, in agreement with the data. We begin by sketching
the most important assumptions that lead to the $Q\propto\sqrt{P}$
prediction of the conventional theory of microwave absorption of TLS;
then we discuss the implications of the much weaker power dependence
reported experimentally in high-Q superconducting CPW resonators.

In the conventional theory, TLS are described by pseudo-spin operators,
$S$, and are characterized by the uniform distribution of the energy
difference, $\varepsilon,$ between their ground and excited state.
In the basis of the eigenstates the Hamiltonian has a simple form
$H=\varepsilon S^{z}$. The ground and the first excited state of
the TLS correspond to the quantum superposition of the states characterized
by different atomic configurations. Each TLS is characterized by a
dipole moment $\hat{\mathbf{p}}=\mathbf{p}(\sin\theta S^{x}+\cos\theta S^{z})$,
which is an operator with both diagonal and off-diagonal components.
$\mathbf{p}$ denotes the difference between the dipole moments in
the two different atomic configurations, its magnitude $p_{0}=|\mathbf{p}|$
sets the scale of the dipole moment. $\theta$ describes the fact
that the eigenstates of the dipole correspond to the superposition
of its states in real space. Because many dipoles have exponentially
small amplitude to tunnel between different positions in real space,
the parameters $\theta$ and $\varepsilon$ are assumed to have distribution
$\mathcal{P}(\varepsilon,\theta)d\varepsilon d\theta\sim\nu/\theta d\varepsilon d\theta$
for small $\theta$, where $\nu=10^{20}/cm^{3}eV$ is the typical
density of states of TLS \cite{Nittke1998}. The interaction between different TLS is
essentially of a dipole-dipole nature with an effective strength given
by the dimensionless parameter $\lambda=\nu p_{0}^{2}/\epsilon$, here
$\epsilon$ is the dielectric constant of the medium, it can also be viewed as coming from TLS coupling to 
elastic strain \cite{Hunklinger1976}. Straightforward
analysis shows that the same parameter also controls the phonon mean
free path at low temperatures \citep{Leggett1991}. The direct measurement
gives values of $\lambda\approx10^{-3}$ in bulk materials, so the
interaction between TLS is usually assumed to be irrelevant.

The intrinsic microwave loss is due to the coupling of the electric
dipole moment $\hat{\mathbf{p}}$ of the TLS to the applied electric
field $\mathbf{E\mathrm{cos\omega t}}$ of the resonator. The resonator
quality factor is related to the imaginary part of the dielectric
function, $\epsilon\left(\omega\right)$, i.e. $Q^{-1}\propto|\Im\epsilon|$.
At microwave frequencies and low temperatures, only the resonant contribution
to the electric susceptibility tensor is relevant. One can compute
the change in the dielectric function due to the resonant response
of an ensemble of TLS by assuming that TLS are relaxed by phonons
and using the Bloch equations that neglect the interaction between
TLS \citep{Hunklinger1976}. In fact, one can get the power dependence
of quality factor $Q$ by following a more general, qualitative argument.
In the steady state driven by sinusoidal electric field, the power
dissipation density is given by the time average 
\begin{equation}
\left\langle P_{d}\right\rangle =\left\langle \mathbf{E\mathrm{\cdot\frac{\partial\mathbf{D}}{\partial t}}}\right\rangle =\frac{\omega}{2}\left|\mathbf{E}\right|^{2}\left|\Im\epsilon\right| \label{eq: Pd}
\end{equation}
 where $\mathbf{D\mathrm{=\epsilon\mathbf{E}}}$ is the displacement
vector. In the absence of TLS-TLS interactions, each TLS can decay
from its excited state only by emitting a phonon; the rate of this
emission gives the decay rate, $\Gamma(\varepsilon)$, of its excited
state. The maximal power that a given TLS can absorb from the photons
in the resonator is $\varepsilon\Gamma$. At small applied powers,
the TLS that absorb photons have energies close to the frequency,
$\omega_{0},$ of the photon field, i.e. $|\varepsilon-\omega_{0}|\lesssim\Gamma$.
It is convenient to characterize the effect of the electric field
on TLS by their Rabi frequency $\Omega=\frac{1}{2}\mathbf{p} \cdot \mathbf{E}\sin\theta$;
for small $\Omega<\Gamma$ TLS at resonance are excited with probability
$(\Omega/\Gamma)^{2}$. Since their density is $\nu\Gamma$, the total
power dissipation density reads $\left\langle P_{d}\right\rangle =\nu\varepsilon\Omega^{2}/\Gamma$.
Different is the situation at larger applied powers. There, even the
TLS with energies further away from the frequency of the resonator,
$\omega_{0},$ get excited. For TLS exactly at resonance, the electric
field $\mathbf{E}\cos\omega_0 t$ results in oscillations with frequency
$\Omega$. When $\Omega>\Gamma$, TLS with energies $|\varepsilon-\omega_{0}|\lesssim\Omega$
are excited. Since their density is $\nu\Omega$, the total dissipated
power density reads $\left\langle P_{d}\right\rangle =\nu\varepsilon\Gamma\Omega$.
By substituting it into Eq. \ref{eq: Pd}, one recovers immediately
the quality factor power dependence $Q\propto\sqrt{\mathrm{\mathrm{P}}}$.

The generality of these qualitative arguments implies that, in the
resonators where a much weaker power dependence of $Q$ has been observed,
the dipoles that absorb the radiation do not get saturated as the
applied power $\mathrm{P}$ is increased. This is possible if
the interactions between dipoles are not negligible and lead to energy
diffusion.

The bulk of evidence indicates that the loss in high-Q superconducting
CPW resonators is due to TLS located on the interface oxide surfaces
(metal-air, metal-substrate or substrate-air) \citep{Wang2009,Lindstrom2009,Macha2010,Wisbey2010,Sage2011},
even though it remains unclear which of the surfaces is more relevant\citep{Wenner2011}.
The estimates show \citep{Pappas2011b} that the concentration of
TLS in these thin surfaces is higher than in the bulk and consequently
the average interaction between TLS is larger, implying that one has
to develop the theory of microwave absorption in $\mathit{interacting}$
TLS in these resonators. In contrast, the resonators that show
$Q\propto\sqrt{\mathrm{\mathrm{P}}}$ and low intrinsic loss in single
photon regime are made of Nb on ${\text{SiO}_{2}/\text{Si}}$ \citep{Lindstrom2009}
and ${\text{AlO}_{x}}$ coated Nb on Si substrate \citep{Pappas2011a} and thus contain 
a significant amount of bulk amorphous dielectrics (${\text{SiO}_{2}}$ or  ${\text{AlO}_{x}}$). It is natural to assume that 
the large intrinsic loss in these resonators is
due to TLS located in the dielectric bulk which is described well
by the conventional theory of independent TLS.

\begin{figure}
\includegraphics[width=3in]{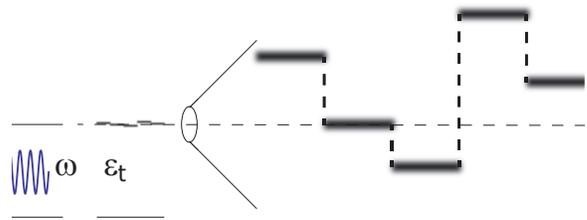}

\caption{Drift of the energy of a given TLS in the field of a few fluctuators.}

\label{Fig:EnergyDrift} 
\end{figure}

Developing the full theory of interacting TLS is a very difficult
problem that was first discussed by Yu et al. \citep{Yu1988} and
still remains controversial \citep{Coppersmith1991,Burin1995b,Lubchenko2001,Legett2011}.
Fortunately, as we shall see below, one does not need to solve the
full theory of interacting TLS in order to estimate the internal loss
of superconducting resonators due to $\mathit{interacting}$ TLS.
In our approach, we assume that the effective degrees of freedom that
remain active in the amorphous oxides surrounding the superconductors
are described by fluctuating dipoles: some of these dipoles are characterized
by fast transitions between their states ($\varepsilon\sin\theta\sim\omega$)
and relatively small decoherence rates, i.e. $\Gamma\ll\omega$, and
are effectively coherent TLS; other dipoles are instead slow and characterized
by decoherence times shorter than the typical time between the transitions, we shall call
them fluctuators. Due to the interaction between fluctuators and TLS,
the frequency of the TLS jumps when the fluctuators in its vicinity
change its state. This translates into the fact that instead of staying
constant the energy of a given TLS drifts with time (see Fig. \ref{Fig:EnergyDrift}).
Notice that this phenomenological model is in a full agreement with
experiments on charge noise performed in superconducting SET and qubits \cite{Kenyon2001}.

We now use this phenomenological model to estimate the internal loss
of the CPW resonators. We begin with qualitative arguments. Due to
the interaction with fluctuators, the TLS with energy level $\varepsilon$
in resonance with the applied electric field stays effectively in
resonance only for a short dwell time $\tau=\gamma^{-1}$ , where
$\gamma$ is the combined relaxation rate of the fluctuators that
affect it (see Fig.\ref{Fig:EnergyDrift}). If the time $\tau$ is
short, i.e. $\tau\Omega\ll1$, the TLS gets into the excited state
with probability $\mathit{\mathit{\mathbb{\mathcal{P_{\mathrm{e}}}}}}\sim(\Omega\tau)^{2}$
and then dissipates the energy away from the resonance. The average
dissipation rate of this process is $\bar{\Gamma}=\gamma\mathcal{P_{\mathrm{e}}\mathrm{\propto\Omega^{2}/\gamma}}$.
Only TLS with energy level located within energy $\gamma$ from the
resonator frequency $\omega$ contribute to this process: their density
is $\nu\gamma$ and the total dissipated power density reads $\left\langle P_{d}\right\rangle =\nu\varepsilon\Omega^{2}\int_{\Omega}^{\gamma_{max}}\mathcal{P}(\gamma)d\gamma$
, where $\mathcal{P}(\gamma)=P_{0}/\gamma$ is the probability distribution
of the fluctuator relaxation rates \cite{Paladino2002}. After integration, we obtain that
$\left\langle P_{d}\right\rangle =\nu P_{0}\varepsilon\Omega^{2}\ln\left(\gamma_{max}/\Omega\right)$
and by inserting it into Eq.\ref{eq: Pd} we find that in this case
$Q\propto\ln\left(\Omega/\gamma_{max}\right)\propto\ln P$, i.e. there
is only a slow, logarithmic dependence of the quality factor with
the applied power $P$.

These qualitative arguments can be confirmed by more quantitative
analytical computations. Before proceeding with the analytic derivation,
we stress that the essential ingredient of the phenomenological model
is the large value of the jumps in the TLS frequency, $\delta\varepsilon_{t}\gg\Gamma$
caused by fluctuators. This implies that the interaction between the
high frequency dipole and the fluctuators in its vicinity is large:
$V(r)=\frac{p_{0}^{2}}{\epsilon r^{3}}\gg\Gamma$. Because the typical
distance between thermally activated TLS is $\bar{r}=\rho^{-1/3}$,
with $\rho=\nu T$, the typical interaction is $V(\bar{r})=\lambda T.$
Comparing it with the relaxation rates of $\Gamma\sim10^{6}s^{-1}$
observed experimentally at frequencies $\omega\sim10$ GHz (and expected
theoretically for phonon emissions) we see that in typical experimental
conditions $\delta\varepsilon\sim\Gamma$ for $\lambda\sim10^{-3}$.
The condition $\delta\varepsilon\gg\Gamma$, however, can be satisfied
for lower frequencies or in materials with larger $\lambda.$ The
estimates of the dissipation per unit volume of surface oxides show
that there their density of TLS is at least 10 times larger than in
the bulk amorphous materials \citep{Pappas2011b}. This translates
into 10 times larger value of the parameter $\lambda$ and a completely
different physics of microwave dissipation described above.

We now give the details for the analytical derivation of the $Q\propto\sqrt{P}$
and its generalization for the model that takes into account energy
drifts of TLS caused by fluctuators. The collection of TLS is characterized
by the Hamiltonian $H_{int}=\sum_{i}\varepsilon_{i}S_{i}^{z}+\frac{1}{2}\sum_{i,j}V_{ab}(r_{i}-r_{j})\mathbf{p}_{a}\mathbf{p}_{b}$$ $.
Here $V_{ab}(r)$ is the interaction between TLS that is due to their
dipole moments or virtual phonon exchange. In either case, in the
static limit, the interaction scales as $V(r)\sim1/r^{3}$. At zero
temperatures and for a small value of the parameter $\lambda\ll1$,
the Hamiltonian $H_{int}$ gives a coherent dynamics of the individual
TLS. In this limit, the levels at high frequencies are broadened by
phonon emission \citep{Black1977} that leads to the relaxation rate
$\Gamma_{1}\propto\varepsilon^{3}$. At non-zero temperatures the
levels are further broadened by dephasing caused by thermal phonons
and other TLS \citep{Faoro2006,Burin1995a}. The combined effect of
all these processes on a single TLS can be described by the Bloch
equations: $\frac{d}{dt}\langle\mathbf{S}(t)\rangle=\left[\langle\mathbf{S}(t)\rangle\times\mathbf{B}(t)\right]-\mathbf{R}(t)$,
where ${\displaystyle \mathbf{R}(t)=\left(\Gamma_{2}\langle S_{x}\rangle,\Gamma_{2}\langle S_{y}\rangle,\Gamma_{1}(\langle S_{z}\rangle-m)\right)}$
is the relaxation matrix, $\mathbf{B}(t)$ is the total field acting
on the TLS and ${\displaystyle m=\frac{1}{2}\tanh\left[\varepsilon/2T\right]}$
denotes the equilibrium population of the TLS levels and $T$ is the temperature. In the conventional
theory of TLS, this field has two components: $\mathbf{B}=\mathbf{B}^{0}+\mathbf{B}^{1}$,
with $\mathbf{B}^{0}=(0,0,\varepsilon)$ and $\mathbf{B}^{1}=(\Omega,0,\Omega')\cos\omega t$
, where $\Omega'=\frac{1}{2}\mathbf{p}\mathcal{\mathbf{E}}\cos\theta$.
The presence of the fluctuators results in the additional time dependence
of the energy levels, $\varepsilon_{t}=\varepsilon+\xi(t)$. Here
$\xi(t)$ denotes a multi-level telegraph noise signal characterized
by switching rate $\gamma.$

The nonlinear Bloch equations are dramatically simplified in realistic
conditions, because the feasible electric field acting on the TLS
has very little effect on them away from the resonance. Formally,
in the stationary solution of the Bloch equations driven with $\mathcal{\mathbf{E}}\cos\omega t$,
all spin components oscillate with frequencies $n\omega$: $\mathbf{S}=\sum_{n}\mathbf{S}_{n}\exp(-in\omega t)$
with $\mathbf{S}_{n}=\mathbf{S}_{-n}^{*}$. Relaxation to the stationary
solution is determined by the decay rate $\Gamma_{1}\ll\Gamma_{2}\ll\omega$
which sets the longest time scale in the problem. It can be described
by allowing slow time dependence of $\mathbf{S}_{n}$ components.
Finally, introducing the operators $S^{\pm}=S^{x}\pm iS^{y}$ and
leaving only the leading resonant terms the Bloch equations reduce
to: 
\begin{align}
i\frac{dS_{1}^{+}}{dt} & =\Omega S_{0}^{z}-(\omega-\varepsilon_{t}+i\Gamma_{2})S_{1}^{+}\label{eq:dS^+/dt}\\
\frac{dS_{0}^{z}}{dt} & =\Omega\Im S_{1}^{+}-\Gamma_{1}(S_{z}^{0}-m)\label{eq:dS^z/dt}
\end{align}

The macroscopic response is given by the average polarization, $\mathbf{P}_{\omega}$
of TLS at frequency $\omega$: 
\begin{equation}
\mathbf{P_{\omega}}=\frac{1}{2}\left\langle \mathbf{p}\sin\theta S_{1}^{+}\right\rangle \label{eq:P_av}
\end{equation}
 where the average is taken over the distribution of TLS and their
dipole moments. The quality factor is directly related to $\mathbf{P}_{\omega}$:
$Q^{-1}\propto\left|\Im\mathbf{P_{\mathrm{\omega}}}\right|/\mathbf{E}$.

If the jumps $\xi(t)$ induced by the fluctuators are small, i.e.
$\xi<\Gamma_{2}$, the stationary solution of Eqs.(\ref{eq:dS^+/dt},\ref{eq:dS^z/dt})
gives the spin component: 
\begin{equation}
S_{1}^{+}=\frac{\Omega(\omega-\varepsilon-i\Gamma_{2})m}{(\omega-\varepsilon)^{2}+\Gamma_{2}^{2}+\Omega^{2}\Gamma_{2}\Gamma_{1}^{-1}}.\label{eq:S^+}
\end{equation}
 At small fields, one can neglect the last term in the denominator
of Eq.(\ref{eq:S^+}). By substituting Eq.(\ref{eq:S^+}) into Eq.(\ref{eq:P_av})
and then averaging over the distribution of TLS, one gets: 
\begin{equation}
|\Im\mathbf{{\bf P_{\omega}}}|=\frac{\pi m}{6}\left\langle \sin^{2}\theta\right\rangle \lambda\mathbf{E}\label{eq:ImP_w_linear}
\end{equation}
 and consequently a $Q$ factor that does not depend on the field
strength \textbf{$\mathbf{E}$}.

At large fields ($\Omega^{2}>\Gamma_{1}\Gamma_{2}$), the third term
in denominator of Eq.(\ref{eq:S^+}) dominates the second and this
results in a rapid decrease of the response. By averaging over the
distribution of $ $TLS, one gets: 
\begin{equation}
|\Im\mathbf{{\bf P_{\omega}}}|=\frac{\pi m}{6}\left\langle \sin^{2}\theta\frac{\sqrt{\Gamma_{1}\Gamma_{2}}}{\Omega}\right\rangle \lambda\mathbf{E}\label{eq:ImP_w_0}
\end{equation}
 which translates into the usual $Q\sim\sqrt{P}$ dependence.

In the opposite case of large jumps $\xi\gg\Gamma_{2}$ one should
solve the full dynamical Eqs.(\ref{eq:dS^+/dt},\ref{eq:dS^z/dt})
with noise $\xi(t)$. The details of the solution depend on the relation
between the relaxation rates $\Gamma_{1},\Gamma_{2}$ and the jump
rate $\gamma$. We discuss first the simplest case of large
$\gamma$ in which a given TLS spends most of its time away from the
resonance, so that when it moves back into the resonance its magnetization
$S_{0}^{z}$ and $S_{1}^{+}$ have their equilibrium values: $S_{0}^{z}=m$
and $S_{1}^{+}=0$. In this case the only quantity that controls the
response is the rate, $\gamma,$ of incoming and outgoing jumps into
the resonance. The stochastic nature of this process implies that,
in order to find the average value of $S_{1}^{+}$ that determines
the response to electric field, we have to solve the equation for
the probability, $\varrho(\mathbf{s})$, to find the resonant TLS characterized
by the three dimensional vector $\mathbf{s}=(x,\, y,\, z)$, where
$x$ and $y$ are the real and the imaginary parts of $S_{1}^{+}$
and $z=S_{0}^{z}$. This probability obeys the evolution equation:
\[
\frac{\partial \varrho}{\partial t}+\frac{d}{d\mathbf{s}}\left(\frac{d\mathbf{s}}{dt}\varrho \right)=\gamma\left[\delta(z-m)\delta(x)\delta(y)-\varrho\right]
\]
 where $d\mathbf{s}/dt$ is given by Eqs.(\ref{eq:dS^+/dt},\ref{eq:dS^z/dt})
with constant $\varepsilon$. The equation is further simplified in
the limit of large $\gamma\gtrsim\Omega\gg\Gamma_{1},\Gamma_{2}$
where the analytic solution gives (see Supplementary material), after
averaging over the distributions of $\varepsilon$ and $\gamma$,
$\mathcal{P}(\varepsilon,\gamma)=\nu P_{0}/\gamma$: 
\begin{equation}
|\Im\mathbf{\mathbf{{\bf P_{\omega}}}|}=\frac{\pi m}{6}\left\langle \sin^{2}\theta\right\rangle \lambda P_{0}\ln\left(\frac{\gamma_{max}}{\Omega}\right)\mathbf{E}\label{eq:ImP_w_1}
\end{equation}
 which results in a weak, logarithmic dependence of the quality factor
$Q$ on the strength of the electric field.

A similar logarithmic dependence occurs in the cases when the dephasing
rate is larger than the jump rates as well. In this case, one can
neglect the time derivative terms in Eq. (\ref{eq:dS^+/dt}) and express
$S_{1}^{+}$ through $S_{0}^{z}$ and get the general evolution equation:
\begin{equation}
\frac{\partial \varrho_{k}}{\partial t}+\frac{\partial}{\partial z}\left[\left(\Gamma_{1}(z-m)-\Xi_{k}z\right)\varrho_{k}\right]=\gamma_{kn}\varrho_{n}\label{eq:p_k_evolution}
\end{equation}
 where $\varrho_{k}(z,t)$ is the probability for a given TLS to have value
$S_{0}^{z}=z$ while subjected to the effective driving field $\Xi_{k}=\Omega^{2}\Gamma_{2}/[(\omega-\varepsilon_{k})^{2}+\Gamma_{2}^{2}]$
in the state $k$ of the fluctuators. The matrix $\gamma_{kl}$ is
the matrix of the transition rates between the states of the fluctuators.
Similarly to the case of the small dephasing rate the saturation of
TLS does not happen if they stay mostly away from the resonance, so
that the probability to find a \emph{given } TLS in resonance $n<\Gamma_{1}/\gamma$
(see Supplementary material). Because the number of states of classical
fluctuators grows exponentially with their number this condition is
satisfied even for a moderate number of fluctuators that affect a given
TLS. In this situation the dissipation is dominated by TLS with $\gamma>\Xi\sim\Omega^{2}/\Gamma_{2}$
: 
\begin{equation}
|\Im\mathbf{P_{\omega}|}=\frac{\pi m}{6}\left\langle \sin^{2}\theta\right\rangle \lambda P_{0}\ln\left(\frac{\gamma_{max}\Gamma_{2}}{\Omega^{2}}\right)\mathbf{E}\label{eq:ImP_w_2}
\end{equation}

Both Eqs(\ref{eq:ImP_w_1},\ref{eq:ImP_w_2}) lead to the weak, logarithmic
dependence of $Q$ at large electric fields ($\Omega^{2}>\Gamma_{2}\Gamma_{1}$),
while at smaller fields, the imaginary part of the average polarization
is given by Eq.(\ref{eq:ImP_w_linear}) and $Q$
is constant.

As a final remark let us notice that in this model the dimensionless coupling between TLS remains small, ${\lambda \ll 1}$. This implies that thermodynamics properties
and the real part of the dielectric constant are not affected by the interaction. Therefore we expect the same temperature-dependent resonance frequency shift
as predicted by the  conventional theory of TLS in agreement with experiments \citep{Anderson1972,Black1977}.

In conclusion, we have shown that the resonance absorption from TLS
is strongly affected by their interaction with classical fluctuators;
this does not change the absorption at low powers but changes the square
root dependence of the absorption into the logarithmic one at high
powers. The important condition is that each fluctuator should move
the TLS frequency more than its width due to the relaxation rate.
This translates into a higher ($\sim 10$ times larger) concentration of TLS than the typical density in amorphous bulk materials..
We interpret the results of recent experiments displaying slow
power dependence of the quality factor in high-Q superconducting CPW
resonators as the evidence for a large concentration of TLS located
in the interface oxide surfaces of the resonator. Very likely it implies that these TLS have a different nature, e.g. localized electron states at
the superconductor oxide boundary. 

{\em Note added}: recent experiment directly observed the energy drift of TLS which is the basis of our model \cite{Grabovskij2012}

The research was supported by ARO W911NF-09-1-0395, DARPA HR0011-09-1-0009
and NIRT ECS-0608842.


\end{document}